\documentclass[aps,pra,epsf,preprint,superscriptaddress,showpacs,preprintnumbers,amsmath,amssymb,11pt]{revtex4-1}
\usepackage{graphicx}
\usepackage{epsfig}
\usepackage{dcolumn}
\usepackage{bm}
\usepackage{braket}
\usepackage{color}
\usepackage{amsmath}
\begin{document}

\title{Interaction Quench Induced Multimode Dynamics of Finite Atomic Ensembles}

\author{S.I. Mistakidis}
\affiliation{Zentrum f\"{u}r Optische Quantentechnologien,
Universit\"{a}t Hamburg, Luruper Chaussee 149, 22761 Hamburg,
Germany}
\author{L. Cao}
\affiliation{Zentrum f\"{u}r Optische Quantentechnologien,
Universit\"{a}t Hamburg, Luruper Chaussee 149, 22761 Hamburg,
Germany}\affiliation{The Hamburg Centre for Ultrafast Imaging,
Universit\"{a}t Hamburg, Luruper Chaussee 149, 22761 Hamburg,
Germany}
\author{P. Schmelcher}
\affiliation{Zentrum f\"{u}r Optische Quantentechnologien,
Universit\"{a}t Hamburg, Luruper Chaussee 149, 22761 Hamburg,
Germany} \affiliation{The Hamburg Centre for Ultrafast Imaging,
Universit\"{a}t Hamburg, Luruper Chaussee 149, 22761 Hamburg,
Germany}

\date{\today}

\begin{abstract}

The correlated non-equilibrium dynamics of few-boson systems in one-dimensional
finite lattices is investigated. Starting from weak interactions we perform 
a sudden interaction quench and employ the numerically exact 
Multi-Layer Multi-Configuration time-dependent Hartree method for
bosons to obtain the resulting quantum dynamics. Focusing on the low-lying modes 
of the finite lattice we observe the emergence of density-wave tunneling, breathing 
and cradle-like processes. In particular, the tunneling induced by
the quench leads to a ''global'' density-wave oscillation. The resulting breathing and cradle modes 
are inherent to the local intrawell dynamics and connected to excited-band states. Moreover, 
the interaction quenches couple the density-wave and the cradle modes allowing for resonance phenomena. 
These are associated with an avoided-crossing in the respective frequency
spectrum and lead to a beating dynamics for the cradle. Finally,
complementing the numerical studies, an effective Hamiltonian in
terms of the relevant Fock states is derived for the description of the
spectral properties and the related resonant dynamics.\\

\end{abstract}

 \maketitle

\section{Introduction}

Ultracold atoms in optical lattices are regarded as an ideal tool to study properties of quantum many-body systems in a controllable manner \cite{Bloch}-\cite{Jaksch}. 
This is experimentally manifested by handling independently the lattice
potential and the interaction strength between the atoms. The former
is achieved by tuning counter-propagating lasers and the latter by
means of optical, magnetic or confinement-induced Fano-Feshbach
resonances \cite{Petsas}-\cite{Olshanii}. 
Currently one of the main focus of many-body physics is to comprehend quantum phase transitions (QPTs) and to unravel their internal mechanisms. 
In this direction, the experimental progress yielded the realization and explanation of Superfluid (SF) to Mott insulating (MI) states, complementing the theoretical efforts within the Bose-Hubbard (BH) framework \cite{Jaksch1}-\cite{Fisher}. 
Furthermore, other exotic quantum phases like the Bose Glass phase or Mott shells have been realized in disordered systems \cite{Batrouni}-\cite{Luhmann}. 
These and other QPTs raise new prospects for theory and experiment, most notably the inescapable necessity of taking quantum effects into account.

Apart from the experimental efforts in the investigation of the ground state properties in many-body systems, recently it became possible, using trapped ultracold atomic gases, to explore the evolution of isolated strongly correlated 
systems \cite{Polkovnikov} after being quenched.
In a corresponding experiment, the system is originally prepared in the ground state $\left| {{\psi _0}} \right\rangle $ of the Hamiltonian ${\hat H_i}$, and then driven out of equilibrium at time $t = 0$ by a sudden change of either the 
trapping frequency or the interaction strength, yielding a new Hamiltonian ${\hat H_f}$ evolving the system in time.
The resulting non-equilibrium situation triggers challenging conceptual questions concerning the unitary evolution, such as the not yet fully understood connection of quantum ergodicity to the integrability of a 
system \cite{Trotzky}-\cite{Rigol}.
The experimental applications in this field includes the realization of a quantum version of Newton's cradle \cite{Kinoshita}, the quenching of a ferromagnetic spinor condensate \cite{Sadler}, the light-cone effect in the spreading 
of correlations \cite{Langen}-\cite{Cheneau}, as well as the collapse and revival of a BEC \cite{Greiner1}.
Also, a recent experiment on quenched atomic superfluids reports the realization of Sakharov oscillations which are known to emerge from the large-scale correlations in galaxies \cite{Hung}.

From another perspective, the inclusion of higher-band contributions results in an additional orbital degree of freedom yielding novel phenomena such as unconventional condensation \cite{Cai}-\cite{Kuklov1} and anisotropic tunneling.
Indeed, excited-band populations caused by interactions have already been observed either by sweeping the magnetic field across a Feshbach resonance \cite{Kohl} or via Raman transitions which couple directly the zeroth band to the first 
excited $p$-band of the lattice \cite{Muller}. 
Other experimental achievements indicate the observation of a 2D superfluid in the $p$-band \cite{Wirth} and the orbital excitation blockade \cite{Bakr} when exciting atoms to higher orbitals as well as supersolid quantum phases in cubic 
lattices \cite{Scarola}-\cite{Scarola1}. 
The aforementioned aspects have led, among others, to the construction of multiflavor and multiorbital models \cite{Isacsson}-\cite{Kuklov}. Motivated by the previous studies here we investigate the higher-band dynamics 
of interaction quenched superfluids focussing on the resulting low-lying collective modes, which are nowadays of great experimental interest.

In the present study, we examine the response of a finite atomic ensemble confined in 1D finite lattices subjected to a sudden change in the
interaction strength. More precisely, we focus on highly non-perturbative situations by considering weak-to-strong interaction quenches with respect to the initial state. 
In this manner, we drive the system to a regime where the interparticle interactions dominate in comparison to their kinetic energy.
For weak interactions, the single-band approximation, namely the BH model, provides quantitative predictions of the system dynamics; 
for strong interactions, however, it yields at most a qualitative description.
In this manner, by considering strong quench amplitudes and examining representative few-body setups for incommensurate filling factors, our treatment goes beyond the validity of the BH model.
The numerical method which we employ in order to study the dynamical properties of our 1D finite setups is the recently developed  Multilayer Multi-Configuration Time-Dependent Hartree method for 
Bosons (ML-MCTDHB)\cite{Cao}-\cite{Kronke}, based on MCTDHB which has been developed and applied successfully previously \cite{Alon}-\cite{Broeckhove}. Both methods are very 
efficient in treating bosonic systems both for static properties and in particular their dynamics (see next section), while they are equivalent for the case of a single species treated here.

We demonstrate the emergence of higher-band modes, namely the breathing and the cradle modes as well as the rise of the density-wave tunneling, following interaction quenches.
Especially the observation of the cradle mode which refers to a localized wave-packet oscillation is arguably one of our central results.
The dynamical properties of incommensurable setups are investigated by examining the time-evolution of the corresponding one-body densities and their respective fluctuations. 
In addition, we analyze the Fourier spectra of representative intrawell observables and the variation of the center of mass coordinate for the cases of the cradle and breathing modes, respectively. 
More specifically, the occurrence of a resonance between the cradle and one of the tunneling modes, being manifested by an avoided crossing in the frequency spectrum, is observed here. This opens the possibility 
to control the interwell dynamics by triggering the intrawell dynamics via the quench amplitude in optical lattices.
Additionally, the construction of an effective Hamiltonian describing the dynamical behaviour is provided and the minimal Fock space required to
produce the cradle process is derived.

The work is organized as follows. 
In Sec. II we introduce our setup, explaining also the ML-MCTDHB method, the quench protocol and the number state representation.
In Sec.~III we report on the quench dynamics for different incommensurate filling factors and demonstrate the emergent modes that arise due to the interaction quench. 
We summarize our findings and give an outlook in Sec.~IV.

\section{Setup and analysis tools. }

\subsection{The Model}
Our system consists of N neutral short-range interacting bosons in a one-dimensional trap. The
many-body Hamiltonian reads
\begin{equation}
\label{eq:1}H = \sum\limits_{i = 1}^N {\left(
{\frac{{p_i^2}}{{2M}} + V({x_i})} \right) + \sum\limits_{i < j}
{{V_{{\mathop{\rm int}} }}(x{}_i - {x_j})} },
\end{equation}
where the one-body part of the Hamiltonian contains the 1D lattice
potential $V({x_i}) = {V_0}{\sin ^2}(k{x_i})$ which is
characterized by its depth ${V_0}$ and periodicity $l$, with $k = \pi
/l$ being the wave vector of the lasers forming the optical
lattice. Furthermore, in order to restrict the infinite trapping
potential $V({x_i})$ to a finite one with $m$ sites and
length $L$, we impose hard wall boundary conditions at the
appropriate positions.
 On the other hand, we model the short range two-body
interaction potential as ${V_{{\mathop{\rm int}} }}({x_i} - {x_j}) =
{g_{1D}}\delta ({x_i} - {x_j})$ with the effective coupling strength
${g_{1D}} = \frac{{2{\hbar ^2}{a_0}}}{{M a_ \bot ^2}}{\left( {1 -
\left| {\zeta  \left( {1/2} \right)} \right|\frac{{{a_0}}}{{{\sqrt{2}a_ \bot
}}}} \right)^{ - 1}}$ \cite{Olshanii}. The coupling $g_{1D}$ depends
on the 3D s-wave scattering length ${a_0}$, the oscillator length ${a_
\bot } = \sqrt {\frac{\hbar }{{M {\omega _ \bot }}}} $ of the
transverse trapping potential and the mass $M$ of the atom. From
the above expression it is obvious that we can tune the interaction
strength by the scattering length ${a_0}$ or the frequency of the
confinement ${\omega _ \bot }$ via Feshbach resonances
\cite{Kohler}-\cite{Chin} or confinement induced resonances
\cite{Kim}-\cite{Giannakeas} respectively. Additionally, for reasons of computational
convenience we will rescale the above Hamiltonian in units of the
recoil energy ${E_R} = {\hbar ^2}{k^2}/2M$ by setting $\hbar  =
M  = 1$. In this manner, the rescaled
interaction strength can be rewritten as $g = \frac{{{g_{1D}}}}{{{E_{R}}}}$, whereas the spatial and temporal coordinates are given 
in units of ${k^{ - 1}}$ and $ E_R^{ - 1}$ respectively. Therefore, all
quantities below are in dimensionless units.

In a BH model which we address here for reasons of comparison, the Hilbert space is truncated 
with respect to the localized lowest-band Wannier states which form a complete set of orthogonal basis functions. This represents an
alternative and more convenient way for discussing phenomena in
which the spatial localization of states plays an important role. Our ab-initio simulation goes beyond the single-band approximation 
and requires higher-band states to describe the
real and site independent (${J_{ij}} = J_{ji}^* \equiv J$) tunneling
strength. Notice also, that the hard wall boundaries we consider here imply
zero tunnel coupling between the first and the last sites (in
contrast to periodic boundary conditions
which result in a certain coupling for all sites). In our ab-initio simulations we use a sufficiently large lattice depth ${V_0}
= 4.5$ such that each well includes two localized single-particle
Wannier states, i.e. the ground and first-excited states, while the
higher excited states are taken into account as delocalized
states.

\subsection{The Computational Method : ML-MCTDHB}

The Multi-Layer Multi-Configuration time-dependent Hartree method for
bosons (ML-MCTDHB) constitutes a variational numerically exact
ab-initio method for investigating both the stationary properties and in particular
the non-equilibrium quantum dynamics of bosonic systems
covering the weak and strong correlation regimes. Its multi-layer
feature enables us to deal with multispecies bosonic systems,
multidimensional or mixed dimensional systems in an efficient
manner. Also, the multiconfigurational expansion of the wavefunction
in the ML-MCTDHB method takes into account higher-band effects which
renders this approach unique for the investigation of systems
governed by temporally varying Hamiltonians, where the system can be
excited to higher bands especially during the dynamics. An important characteristic of
the ML-MCTDHB approach is the representation of the wavefunction by variationally
optimal (time-dependent) single particle functions (SPFs) and expansion
coefficients ${A_{{i_1}...{i_S}}}(t)$. This renders the
truncation of the Hilbert space 
optimal when employing the optimal time-dependent moving basis. Also, the requirement for convergence demands a sufficient number of SPFs such that the numerical 
exactness of the method is guaranteed. Therefore, the number of SPFs has to be increased until the quantities of interest acquire the corresponding numerical accuracy. This constitutes a numerically 
challenging and time-consuming task especially for strong interactions where the use of more SPFs to ensure convergence is unavoidable.

Let us elaborate. In a generic mixture system consisting of ${N_\sigma }$ bosons of
species $\sigma  = 1,2,...,S$ the main concept of the ML-MCTDHB
method is to solve the time-dependent Schr\"{o}dinger equation
\begin{equation}
\label{eq:2}\begin{array}{l}
i {\ket{\dot{\Psi}}  }   = \widehat {\rm H}\left| \Psi  \right\rangle\\

\ket{{\Psi (0)}}  = \left| {{\Psi _0}} \right\rangle,
\end{array}
\end{equation}
as an initial value problem by expanding the total wave-function in
terms of Hartree products
\begin{equation}
\label{eq:3}\left| {\Psi (t)} \right\rangle  = \sum\limits_{{i_1} =
1}^{{M_1}} {\sum\limits_{{i_2} = 1}^{{M_2}} {...\sum\limits_{{i_S} =
1}^{{M_S}} {{A_{{i_1}...i{}_S}}(t)} } } \ket{{\psi
_{{i_1}}^{(1)}(t)}} ... \ket{{\psi _{{i_S}}^{(S)}(t)}}.
\end{equation}
Here each species state $\ket{\psi  _i^{(\sigma )}}$ ($i =
1,2,...,{M_\sigma }$) corresponds to a system of ${N_\sigma }$
indistinguishable bosons, which in turn can be expanded in terms of
bosonic number states ${\ket{{\vec n (t)}}^\sigma }$ as follows
\begin{equation}
\label{eq:10}\ket{\psi  _i^{(\sigma )}}  = \sum\limits_{\vec n
\parallel \sigma } {C_{i;\vec n }^\sigma (t)} {\left| {\vec n (t)}
\right\rangle ^\sigma },
\end{equation}
where each $\sigma $ boson can occupy ${m_\sigma }$ time-dependent
SPFs $\ket{\varphi _j^{(\sigma )}}$.
The vector $\left| {\vec n } \right\rangle  = \left|
{{n_1},{n_2},...,{n_{{m_\sigma }}}} \right\rangle$ contains the
occupation number ${n_j}$ of the $j - th$ SPF that obeys the
constraint ${n_1} + {n_2} + ... + {n_{{m_\sigma
}}} = {N_\sigma }$.

Here we focus on the case of a single species in one-dimension where the ML-MCTDHB is equivalent to MCTDHB \cite{Alon}-\cite{Broeckhove},\cite{Beck}. To be self-contained, let us
briefly discuss the ansatz for the many-body wavefunction and the
procedure for the derivation of the equations of motion. The many-body wavefunction 
is a linear combination of time-dependent permanents
\begin{equation}
\label{eq:10}\left| {\Psi (t)} \right\rangle  = \sum\limits_{\vec n
} {{C_{\vec n }}(t)\left| {{n_1},{n_2},...,{n_M};t} \right\rangle },
\end{equation}
where $M$ is the number of SPFs and the
summation is again over all possible combinations which retain the
total number of bosons. Notice that in the limit in which M
approaches the number of grid points the above expansion becomes
exact in the sense of a full configuration interaction approach. On the other hand, the permanents in (5) can be expanded in
terms of the creation operators $a_j^\dag (t)$ for the $j - th$
orbital ${\varphi _j}(t)$ as follows
\begin{equation}
\label{eq:4}\left| {{n_1},{n_2},...,{n_M};t} \right\rangle  =
\frac{1}{{\sqrt {{n_1}!{n_2}!...{n_M}!} }}{\left( {a_1^\dag }
\right)^{{n_1}}}{\left( {a_2^\dag } \right)^{{n_2}}}...{\left(
{a_M^\dag } \right)^{{n_M}}}\left| {vac} \right\rangle,
\end{equation}
which satisfy the standard bosonic commutation relations $\left[
{{a_i}(t),{a_j}(t)} \right] = {\delta _{ij}}$, etc. To proceed
further, i.e. to determine the time-dependent wave function $\left|
\Psi \right\rangle$, we have to find the equations of motion for the
coefficients ${{C_{\vec n }}(t)}$ and the orbitals (which are both
time-dependent). For that purpose one can employ various schemes
such as the Lagrangian, McLachlan \cite{McLachlan} or the
Dirac-Frenkel \cite{Frenkel}-\cite{Dirac} variational principle, each
of them leading to the same result. Following the Dirac-Frenkel
variational principle
\begin{equation}
\label{eq:5}{\bra{\delta \Psi}}{i{\partial _t} - \hat{ H}\ket{\Psi
}}=0,
\end{equation}
we can determine the time evolution of all the coefficients
${{C_{\vec n }}(t)}$ in the ansatz (5) and the time dependence of
the orbitals $\left| {{\varphi _j}} \right\rangle $. These appear as a  coupled system of ordinary differential equations for the time-dependent 
coefficients  $C_{\vec{n}}(t)$ and non-linear integrodifferential equations for the time-dependent 
orbitals $\phi_{j}(t)$. The aforementioned equations constitute the well-known MCTDHB equations of motion \cite{Alon}-\cite{Broeckhove}.

Note that for the needs of our implementation we have used a
discrete variable representation for the SPFs (or orbitals)
$\left| {{\varphi _j}} \right\rangle$, specifically a sin-DVR which
intrinsically introduces hard-wall boundaries at both ends of the
potential (i.e. zero value of the wave function on the first and the
last grid point). For the preparation of our initial state
we therefore relax the bosonic wavefunction in the ground state of the
corresponding $m$-well setup via imaginary time propagation in the
framework of ML-MCTDHB. Subsequently, we change abruptly the 
interaction strength and explore the time evolution of $\Psi
({x_1},{x_2},..,{x_N};t)$ using ML-MCTDHB. Finally, note that in order to justify the convergence of our simulations, e.g. for the triple well, we have used up to 10 single particle functions finally confirming 
the convergence.  Another criterion for convergence is the population of the natural orbital with the lowest population which is kept for each case below $0.1\%$.

\subsection{Quantum quench protocol}
 
Our approach to study the nonequilibrium dynamics follows a
so-called quantum quench. According to this the system is originally
prepared at $t = 0$ in the ground state $\left| {{\psi _0}}
\right\rangle$ of some initial Hamiltonian ${H_{in}} = H({\zeta
_{in}})$, where ${\zeta _{in}}$ is a system parameter associated to
the perturbation such as the interaction strength or the height of the
barrier. Then for times $t> 0$ we suddenly quench the parameter
$\zeta$ to a final value ${\zeta _f}$ and examine the subsequent
evolution of the system under the new Hamiltonian ${H_f} = H({\zeta _f})$.

In the general case, the final Hamiltonian assumes the form ${H_f} =
{H_{in}} + \lambda {H_r}$, where ${H_r}$ is a dimensionless
perturbing operator and $\lambda $, which possesses the dimensionality 
of an energy, is the so-called quench amplitude. In our case the quench protocol
consists of tuning the interaction strength between the particles
which appears in the two-body part ($V_{int}$) of the Hamiltonian
(1). Therefore, we assume as the initial state $\left| {{\psi _0}}
\right\rangle $ (at $t = 0$) the ground state of the Hamiltonian
${H_{in}} = H({g_{in}})$ and we explore its dynamical behaviour
for $t> 0$ subject to the Hamiltonian ${H_f} = H({g_f})$. Under
this protocol the time evolution of the system according to the
Schr\"{o}dinger picture is $\left| {\psi (t)} \right\rangle  = {e^{
- \frac{i}{\hbar }{H_f}t}}\left| {{\psi _0}} \right\rangle $ while
the evolution of the expectation value of a system operator $\hat A$
obeys
\begin{equation}
\label{eq:6}\left\langle {\psi (t)} \right|\widehat {\rm A}\left|
{\psi (t)} \right\rangle  = \sum\limits_{f,f'} {{C_f}{C_{f'}}{e^{ -
i\frac{{{E_f} - {E_{f'}}}}{\hbar }t}}} \left\langle f \right|\widehat
A\left| {f'} \right\rangle,
\end{equation}
where $\left| f \right\rangle$ refers to the eigenstates and ${E_f}$
the respective eigenvalues of the final Hamiltonian ${H_f} =
H({g_f})$. Thus, for our system the new Hamiltonian governing the
dynamics can be written as follows
\begin{equation}
H\left( {{g_f}} \right) = H\left( {{g_{in}}} \right) +
\frac{{\delta g}}{{{g_{in}}}}\sum\limits_{k < j} {{V_{{\mathop{\rm
int}} }}({x_k} - {x_j})},
\end{equation}
with $\frac{{\delta g}}{{{g_{in}}}}$ being the corresponding
quench amplitude.

\subsection{Number state representation}

Using ML-MCTDHB we calculate the wavefunction with respect to a
time-dependent basis of SPFs. Therefore the expansion of the wavefunction in general reads $\left| {\psi (t)} \right\rangle
= \sum\limits_{\vec n } {{A_{\vec n }}(t)\left| {n(t)} \right\rangle
}$. On the analysis side, however, it is preferable to analyze our results in a time-independent basis and
make the connection with the multiband Wannier functions. In this respect,
we have developed in the framework of ML-MCTDHB a fixed basis
analysis package in terms of which we use a time-independent basis for the
expansion of the wavefunction, i.e. $\left| {\psi (t)} \right\rangle
= \sum\limits_{\vec m } {{{\widetilde {\rm A}}_{\vec m }}(t)\left|
{\vec m } \right\rangle } $.

In addition, in order to interpret our results we will use as an
explanatory tool the concept of a generalized number state
representation with multiband Wannier states. To use this representation we assume that the lattice potential is deep 
enough such that the Wannier functions belonging to different wells have very small overlap for not too high 
energetic excitation. Within this
framework we can analyze the interband processes as well as the
intraband tunneling. As a specific example, let us elaborate for the
case of a triple well system the corresponding wavefunction in terms
of these states which encode the allocation of the $n$ bosons among
the individual wells
\begin{equation}
\label{eq:7}\left| \psi  \right\rangle  = \sum\limits_{n,I}
{{C_{n;I}}{{\left| {{n_L},{n_M},{n_R}} \right\rangle }_{I}}}.
\end{equation}
Here ${{n_L},{n_M},{n_R}}$ are the number of bosons localized in the
left, middle, and right well respectively which satisfy the
condition ${n_L} + {n_M} + {n_R} = n$. The summation is over all the
different arrangements of the $n$ bosons in the triple well as well
as the different necessary excited states (index $I$) that we must
take into account according to their energetical order. In this manner,
we use an expansion in terms of the number states of the
non-interacting bosons, i.e. products of non-interacting single particle
Wannier functions. Finally, it is important to notice that such an expansion is valid also in the
strong interaction regime but needs then a large number of excited configurations.

The notion of the generalized number states will be one of our basic
tools for the analysis of the non-equilibrium dynamics. For illustration, let us
analyze in some detail the case of four bosons in a triple well
which will be one of the considered setups in the following.
Here, in terms of the number states we can realize
four different categories. The quadruple mode $\{ {\left| {4,0,0}
\right\rangle _I},{\left| {0,4,0} \right\rangle _I},{\left| {0,0,4}
\right\rangle _I}\} $ that refers to four bosons in the same well,
as well as the triple mode $\{ {\left| {3,1,0} \right\rangle _I},{\left|
{0,3,1} \right\rangle _I},{\left| {1,0,3} \right\rangle _I},{\left|
{1,3,0} \right\rangle _I},{\left| {0,1,3} \right\rangle _I},{\left|
{3,0,1} \right\rangle _I}\}$ which implies that three bosons are localized
in the same well and the fourth resides in one of the remaining wells. In
addition, there is the pair mode that can be separated into two categories: the
double pair mode $\{ {\left| {2,2,0} \right\rangle _I},{\left|
{0,2,2} \right\rangle _I},{\left| {2,0,2} \right\rangle _I}\}$ where
the bosons are divided into two pairs each of them occupying a
different well and the single pair mode $\{ {\left| {2,1,1}
\right\rangle _I},{\left| {1,2,1} \right\rangle _I},{\left| {1,1,2}
\right\rangle _I}\}$ which contains a pair and two separated
bosons. 

Let us comment on the relation between 
the different categories of number states and the eigenstates of the 
system. The number states of a particular category with the same intrawell energetical index $i$ 
share a similar ''on-site'' energy and they will significantly contribute to the same eigenstates. In this manner, 
the eigenstates can be also classified with respect to the dominantly contributing number states, e.g. the 
single-pair (SP), the double-pair (DP), the triple (T) and the quadruple (Q) mode. To be concrete, in the following we 
will use the notation ${\left|{i} \right\rangle _{\alpha;I}}$ to characterize the eigenstates, where the index $\alpha$ refers to the spatial occupation, 
i.e. the SP ($\alpha=1$), DP ($\alpha=2$), T ($\alpha=3$) and Q ($\alpha=4$), the index $I$ denotes the respective energetical level and $i$ stands for the 
index within each group. For instance ${\{\left|{i} \right\rangle _{1;0}}\}$ represent the eigenstates which are dominated by the set of single pair states $\{ {\left| {2,1,1}
\right\rangle _0},{\left| {1,2,1} \right\rangle _0},{\left| {1,1,2} \right\rangle _0}\}$, where the index $i$ take values from 1 to 3.

Finally, note that for the second system which we consider,
consisting of five bosons in ten wells, the same analysis in terms of number
states is straightforward. More precisely, one can realize seven
different categories of number states. Namely, the single mode
$\{ {\left| {1,1,1,1,1,0,...} \right\rangle _I},...\}$, the single pair mode
$\{ {\left| {2,1,1,1,0,...} \right\rangle _I},...\}$, the double pair mode $\{ {\left| {2,2,1,0,...} \right\rangle _I},...\}$, 
the first triple mode $\{ {\left| {3,1,1,0,...} \right\rangle _I},...\}$, the second triple mode $\{ {\left| {3,2,0,...} \right\rangle _I},...\}$ 
the quadruple mode $\{ {\left| {4,1,0,...} \right\rangle _I},...\}$ and the fifth mode $\{ {\left| {5,0,...} \right\rangle _I},...\}$. Here, each  mode can be 
characterized using similar arguments as we did for the case of the triple-well.

\section{Quench dynamics}
The main characteristic of a system with incommensurate filling $\nu$ is
the existence of a delocalized fraction of particles \cite{Brouzos}.
Therefore, the most important feature is the absence
of a Mott insulating state since there is a superfluid fraction on
top of a Mott insulator phase. Below we consider both the case $\nu
> 1$ where on-site interaction effects prevail and $\nu  < 1$ in
which the main concern is the redistribution of particles over
the sites as the interaction increases. In the following, we proceed
for each case with a brief discussion of the ground state properties
and then we focus on the quantum dynamics resulting after an
interaction quench.

\subsection{Filling factor $\nu  > 1$}

Our initial state is the ground state for a given interaction strength in the
weak-interaction regime. Therefore, let us briefly summarize
the ground state properties for weak interactions for a setup
consisting of four atoms in three wells, i.e. one extra particle on
a Mott background. For this case of
incommensurability we encounter two main
aspects: delocalization and on-site interaction effects. The
particle density for the non-interacting case $g = 0$ is largest in the middle site and decreases for the outer ones due to the hard-wall 
boundary conditions that render the middle and outer sites non-equivalent. In the low-interaction
regime we observe a tendency towards a uniform population for $g \approx
0.2$ due to the repulsion of the bosons. For further increasing repulsion such as $g = 0.8$ we note a trend
towards the repopulation of the central well again.

\begin{figure}[h]
        \centering
           \includegraphics[width=0.90\textwidth]{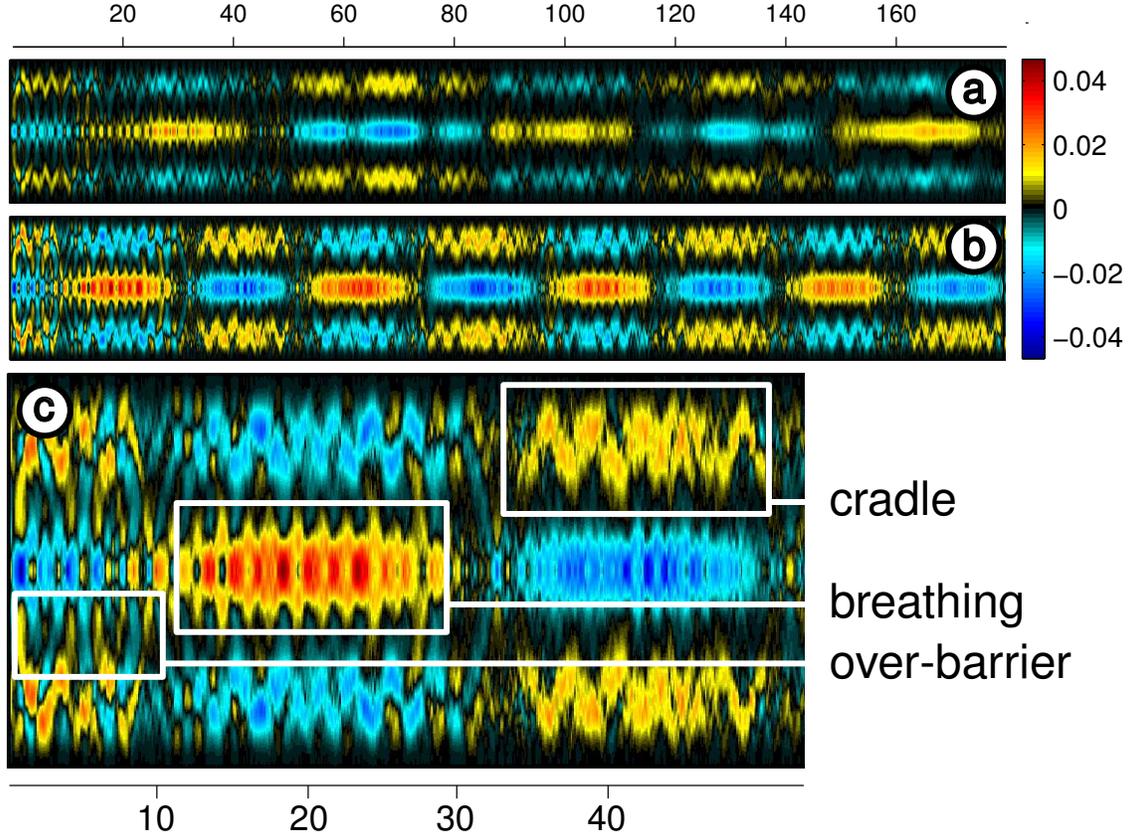}
                \caption{The fluctuations $\delta\rho(x,t)$ of the one-body density caused by an abrupt quench of the inter-particle
                repulsion. The initial state of each setup is the ground state of N=4 bosons confined in a triple-well trap with
                $g_{in}=0.05$. The space-time evolutions of the density are depicted for different quench amplitudes (a)
                $\delta g = 0.8$, (b) $\delta g = 2.0$. In (c) we show an inset of (b) for the first $t=50$ time units where we  
                demonstrate the cradle, breathing and over-barrier modes. Note that the spatial extent of each well is (-$3\pi/2$:-$\pi/2$), (-$\pi/2$:$\pi/2$), 
                ($\pi/2$:$3\pi/2$) for the left, middle and right wells respectively. The vertical axis represents the spatial extent of the trap whereas 
                the horizontal axis denotes the propagation time $t$.}
\end{figure}
In the following, we study the quench dynamics for $t > 0$ of the
above setup by means of an abrupt change in the repulsive
interaction strength at $t = 0$.
\begin{figure}[h]
        \centering
                \includegraphics[width=0.90\textwidth]{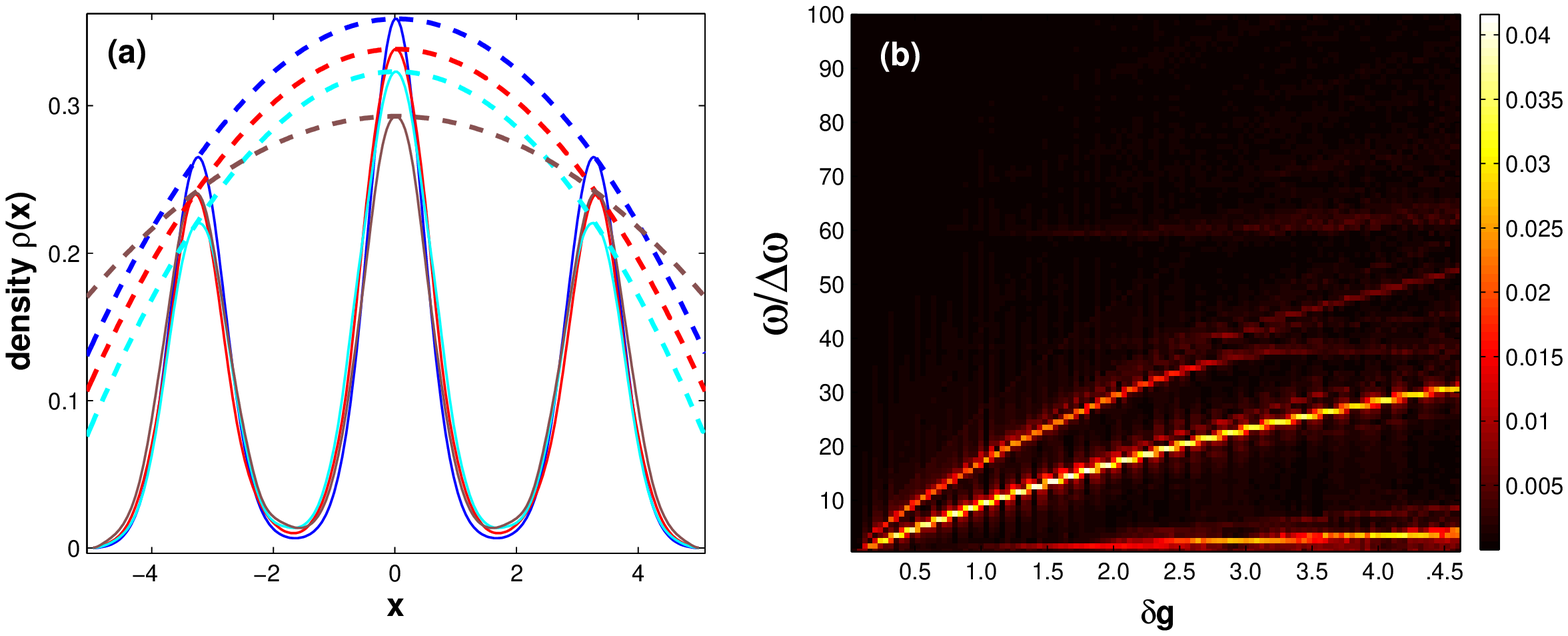}
                \caption{ (a) Evolution of the one-body density
                $\rho(x,t)$ induced by an abrupt change of the inter-particle repulsion with amplitude $\delta{g}=2.8$. 
                The initial state is the superfluid ground state of N=4 bosons with $g_{in}=0.05$ confined in a triple-well trap.
                We observe spatio temporal oscillations constituting the density waves (see also Figure
                1). Shown is also the envelope of the one-body density (dashed lines) at different time instants: t=1 (blue), t=10.3 (red), t=18.8 (light blue) and t=26.2 (brown). The spectrum of the interwell tunneling
                 modes can be obtained from the spectrum of the fidelity $F(t) = {\left| {\left\langle {\psi (0)} \right|\left. {\psi (t)} \right\rangle } \right|^2}$
                 which is shown in (b) as a function of the quench amplitude $\delta{g}$. Here the vertical axis refers to normalized frequency units $\omega/\Delta\omega$, where 
                 $\Delta\omega=2\pi/T$ and T being the respective propagation time.}
\end{figure}
In order to investigate out-of-equilibrium aspects in our system we
first examine the response of the one-body density. Therefore, we perturb our system starting from a
superfluid ground state with ${g_{in}} = 0.05$ where the atoms are
bunching in the central well. As a consequence of the quench the
system gains energy. Figures 1(a)-(b)  show the
time-evolution of the relative density in the triple well trap 
for weak and strong quench amplitudes, namely
$\delta g = 0.8$ and $\delta g = 2.0$ respectively. We define the
deviation of the instantaneous density from the average value up to time $T$ for each grid point x as ${\delta\rho} (x,t)
= {\rho}(x,t) - {\left\langle {{\rho(x)}} \right\rangle _T}$ where
the quantity $\left\langle {{{\left. \rho(x) \right\rangle }_T} = }
\right.{{\int\limits_0^T {dt} {\rho}(x,t)} \mathord{\left/
 {\vphantom {{\int\limits_0^t {dt} {\rho _i}(t)} T}} \right.
 \kern-\nulldelimiterspace} T}$ refers to the corresponding mean single-particle probability density. Therefore, ${\left\langle
{{\rho(x)}} \right\rangle _T}$ refers to the average behaviour of the
one-body density while $\delta\rho(x,t)$ is
the respective fluctuating part. According to our simulations the
ratio $\frac{\left|\delta\rho(x,t)\right|}{{\left\langle {{\rho(x)}} \right\rangle _T}}$ is of the order of ${10^{ - 1}}$.

As can be seen in Figure 1,  at each time instant $\delta\rho(x,t)$ exhibits a density-wave like spatial pattern. This density wave also evolves in time, 
changing between a peak-valley-peak and a valley-peak-valley pattern, where the peak and valley refer to a positive and negative relative density in a certain well, respectively. The evolution of this 
pattern reflects the tunneling dynamics under a quench, and will be termed in the following as density-wave tunneling. Note that the density-wave tunneling 
refers exclusively to the mode that transfers population 
among the middle and the outer wells. Additionally the inner-well dynamics which can be seen 
in Figure 1(c) is described by two excited modes: the middle well exhibits
a breathing mode due to the lattice symmetry, while in the left and
right wells we observe the so-called cradle mode, manifested as a "dipole-like" oscillation of the localized
wavepacket. A close comparison of Figures 1(a) and 1(b) reveals a transition from a
multifrequency to a single frequency spectrum for
weak to strong interaction quenches respectively.
\begin{figure}[h]
        \centering
                \includegraphics[width=0.90\textwidth]{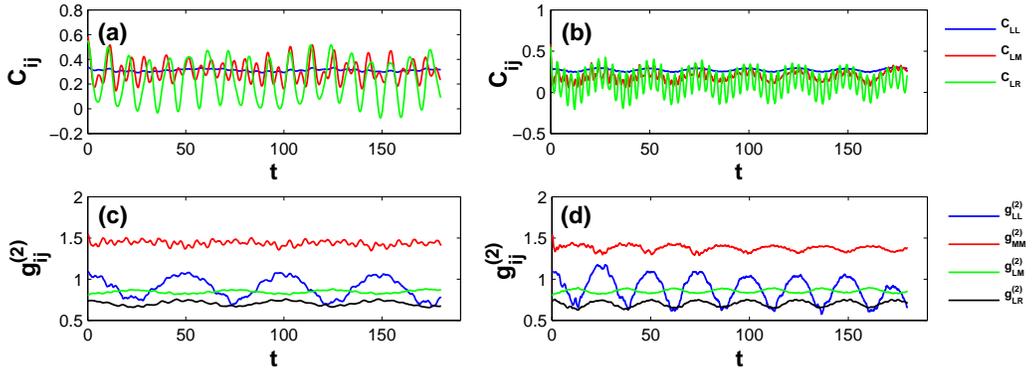}
                \caption{ The time evolution of the one-body
                correlation function $C_{ij}$ and the density
                correlations $g^{(2)}_{ij}$ for a
                quench from $g_{in}=0.05$ to (a) $g_{f}=1.0$ and (b) $g_{f}=4.2$ (see text).
                For the density correlations we demonstrate the situation of
                (c) a weak quench $\delta{g}= 0.6$ and (d) a strong
                quench $\delta{g}= 4.0$.}
\end{figure}
In the following, we will discuss in some detail each of the
aforementioned dynamical modes and their significant role in the
overall nonequilibrium dynamics.

\subsubsection{Density-wave tunneling and breathing mode}

Let us first focus on the explanation of the density-wave 
tunneling as an effective breathing of the "global
wavepacket" described by the envelope of the density distribution in
the triple well. According to this, we illustrate in Figure 2(a) some
intersections of the one-body density for different time instants
and define an envelope function for the triple-well which
is the quadratic function that encloses the corresponding
instantaneous peaks of the density. As we have already mentioned, the density-wave reflects the tunneling dynamics of bosons confined in the optical lattices, which is dominated in the
present case by the states of the lowest-band. In turn, the dynamical tunneling is constituted by the contraction and expansion of the envelope in the course of the dynamics induced by the interaction quench. Intuitively, under 
an interaction quench the bosons tend to repel each other and the envelope will expand and
then contract, which mimics the breathing dynamics of the bosons as known in
the harmonic trap. In a recent study \cite{Tschischik} this mode has been 
examined in the framework of the BH model for a quench in the lattice
frequency. This suggests that it also exists for
many-body systems in optical lattices where instead of our
hard-wall boundaries a weak harmonic confinement renders the sites of the optical lattice non-equivalent.
\begin{figure}[h]
        \centering
                \includegraphics[width=0.90\textwidth]{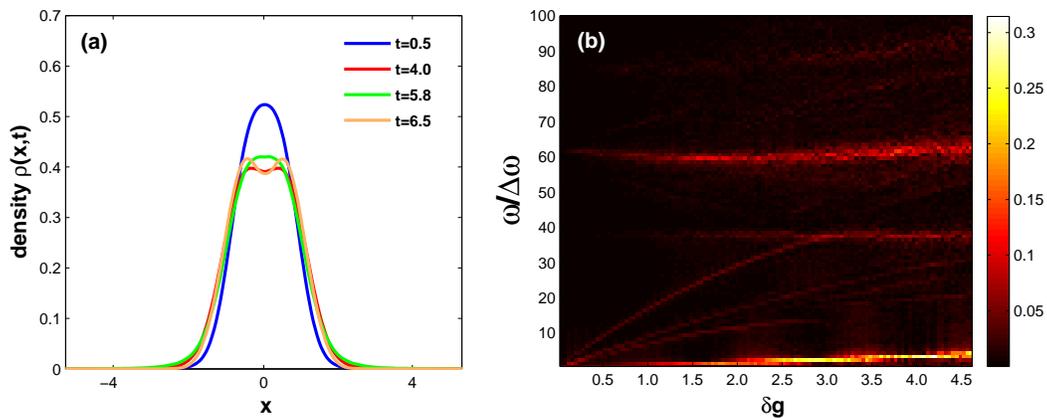}
                \caption{ (a) One-body density profiles at different
                time instants after an interaction quench. The
                system consists of N=2 bosons confined in a harmonic
                trap with $g_{in}=0.05$, while the quench amplitude
                is $\delta{g}=2.8$. The reshaping of the density
                indicates the breathing mode while the oscillatory
                structure demonstrates the contribution of excited
                states during the dynamics. On the other hand, in (b) we present the Fourier spectrum as a function of the quench amplitude for the quantity
                 $\sigma _M^2(\omega )$ referring to the breathing mode. We observe
                that the breathing frequency is predominantly constrained to a narrow band. Note that we use normalized frequency units $\omega/\Delta\omega$, with 
                 $\Delta\omega=2\pi/T$ and T being the whole propagation time.}
\end{figure}

Let us further investigate the properties of the
tunneling modes due to their significance for the above-discussed effects. 
The tunneling properties can be identified in terms
of the overlap of the instantaneous wavefunction during the dynamics
and the initial state (see eq.(11) below) which we denote as
$D(t) = \left\langle {\psi (0)} \right|\left. {\psi (t)}
\right\rangle $. Then, the quantity that we are interested in
is the probability that the states of the unperturbed and perturbed
system are the same during the time evolution which can be expressed
through the fidelity $F(t) = {\left| {D(t)} \right|^2}$. The
identification of the interwell tunneling branches can be achieved
via the frequency spectrum of the fidelity $F(\omega ) =
\frac{1}{\pi }\int {dt} F(t){e^{i\omega t}}$ which provides us with
the evolution of the frequencies of the tunneling modes for different
quench amplitudes. Figure 2(b) therefore shows
$F(\omega )$ with varying quench amplitude
where we can identify three interwell tunneling branches. Note that
the lowest one dominates for strong interaction quenches and this
can be linked to the transition from a multifrequency to a single
frequency behaviour that we have observed above in Figures 1(a) and 1(b).

Next, in order to obtain a quantitative description of the
multiband behaviour we adapt the number state basis (Section II-D)
where the four different categories consist of: the quadruple,
the triple, the double pair and the single pair mode. Indeed, let
$\left|{\psi (0)} \right\rangle  = \sum\limits_{i;\alpha;I}
{{C_{i}^{\alpha;I}}\left| i \right\rangle _{\alpha;I}}$ be the initial
wavefunction in terms of the eigenstates ${\left| i
\right\rangle }_{\alpha;I}$ of the final Hamiltonian. Then the fidelity reads
\begin{equation}
\label{eq:8}{\left| {\left\langle {\psi (0)} \right|\left. {\psi
(t)} \right\rangle } \right|^2} = \sum\limits_{i;\alpha;I } {{{\left|
{{C_{i}^{\alpha;I}}} \right|}^4} + \sum\limits_{i,j;\alpha ,\beta;I }
{{{\left| {{C_{i}^{\alpha;I}}} \right|}^2}{{\left| {{C_{j}^{\beta;I}}}
\right|}^2}\cos ({\epsilon _{i}^{\alpha;I}} - {\epsilon _{j
}^{\beta;I}})t} },
\end{equation}
where the indices $\alpha ,\beta $ specify the
particular groups of number states introduced in Section II-D, 
$i,j$ is the internal index within each group and $I$ denotes the band index. 
For the density-wave mode that we examine here we have $I=0$. Moreover, in the above expansion 
the terms of the second sum   
represent the different tunneling branches whose Fourier transforms are shown in Figure 2(b). The
eigenstate ${\left| i\right\rangle }_{\alpha;I}$ may belong to one of the
four existing categories of number states with a
corresponding eigenenergy. In particular, the lowest branch in the Fourier spectrum 
corresponds to the energy difference $\Delta\epsilon$ within the energetically lowest 
states of the single pair mode, i.e. intraband tunneling from the
state ${\left| {1,2,1} \right\rangle _0}$ to ${\left| {2,1,1}
\right\rangle _0}$ etc. The second branch refers to the next energetically 
closest different modes. The tunneling
process is here from the energetically lowest state of a single pair mode to the
energetically lowest double pair mode, e.g. from ${\left| {1,2,1} \right\rangle
_0}$ to ${\left| {2,2,0} \right\rangle _0}$. Finally the third
branch refers to a tunneling process from a single pair mode to a 
triple mode, e.g. from ${\left| {2,1,1}
\right\rangle _0}$ to ${\left| {3,1,0} \right\rangle _0}$. The
remaining tunneling branches as for instance a transition from a
double pair mode to a triple mode do in principle exist but they
are negligible in comparison to the above ones due to the respective
energy differences and therefore we can hardly identify them in 
Figure 2(b). Note that the same spectrum could also be found from
the frequency spectrum of the local density of a certain well, 
e.g. from ${\rho _L}(\omega )$.

According to the above the tunneling dynamics here is mainly an
intraband phenomenon. To verify this we have also employed the respective single-band BH model where we have
identified each branch in the weak interaction regime.
Within this framework, we can observe the interwell tunneling processes but have to restrict ourselves to
the weak interaction regime where the single-band approximation is
valid. On the contrary, we can not observe either of the on-site breathing or
cradle motion (see next section) which include higher-band
contributions and are intrinsically linked
to the intrawell structure.

 Another important tool in order to explore the interwell tunneling is to examine how correlations among
different sites react after an interaction quench. We examine two
different types of correlations, the single particle correlations
${C_{ij}}(t) = {{\left\langle \psi \right|b_i^\dag {b_j}\left| \psi
\right\rangle } \mathord{\left/
 {\vphantom {{\left\langle \psi  \right|b_i^\dag {b_j}\left| \psi  \right\rangle } N}} \right.
 \kern-\nulldelimiterspace} N}$ and the second order normalized
correlation function (or coherence) $g_{ij}^{(2)}(t) =
{{\left\langle {\left. {{n_i}{n_j}} \right\rangle } \right.}
\mathord{\left/
 {\vphantom {{\left\langle {\left. {{n_i}{n_j}} \right\rangle } \right.} {\sqrt {\left\langle {\left. {{n_i}} \right\rangle \left\langle {\left. {{n_j}} \right\rangle } \right.} \right.} }}} \right.
 \kern-\nulldelimiterspace} {\left\langle {\left. {{n_i}} \right\rangle \left\langle {\left. {{n_j}} \right\rangle } \right.} \right. }}$. Here, $b_i^\dag$ ($b_i$) denotes the corresponding creation (annihilation) operator of a particle located at site $i$ in the lowest-band, while ${{n_i}}=b_i^\dag {b_i}$ is the number operator 
for the site $i$. Notice that we mainly focus on the lowest-band description as the present tunneling mode is dominated by the lowest-band contributions, thus filtering out the influence from higher-bands. In Figures 3(a), (b) we illustrate the
time evolution for the various types of one-body correlations
associated with the left well for
 different quench amplitudes $\delta g=0.95$ and $\delta g=4.15$ each time starting from the
 superfluid regime (${g_{in}} = 0.05$). The single particle correlations
 oscillate even for long time scales which can be attributed to the
 finite-size of our system. The diagonal term ${C_{LL}}$ reflects the density oscillations of the left
 well which are relatively small. Moreover, we observe the change in the periods T
 of the tunneling, that is as we increase the interaction quench we
 obtain a decrease of the respective period denoted by rapid small amplitude oscillations.
 However, the non-diagonal terms ${C_{ij}}$ with $i \ne j$
 exhibit a non-vanishing oscillatory behaviour with an amplitude much larger than 
the density oscillations, i.e. $C_{LL}$. The latter shows more frequencies than the density which
 illustrates the emergence of more dynamical structures. This indicates
 that even a weak tunneling can transport substantial off-site correlations in the
 system. 
\begin{figure}[h]
        \centering
                \includegraphics[width=0.90\textwidth]{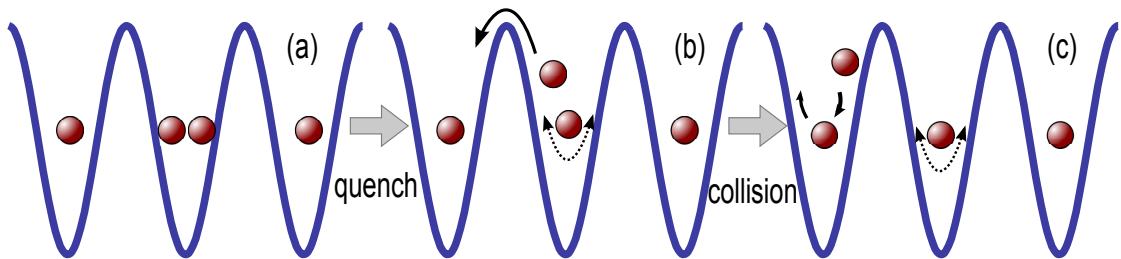}
                \caption{Visualization of the cradle process induced by the over-barrier transport. In this
                scenario, the system which is (a) initially in a
                superfluid ground state is subjected to an abrupt
                interaction quench. In this manner, a boson initiated in the middle well can overcome the barrier (b)
                and move to the neighboring well resulting in a cradle motion (c) due to the quench in the inter-particle
                repulsion.}
\end{figure}

On the other hand, the two-body correlation function $g_{ij}^{(2)}$
can be used to measure density fluctuations in the system under
consideration. A basic property of this function is that
$g_{ij}^{(2)} > 1$ refers to bunching whereas
$g_{ij}^{(2)} < 1$ indicates antibunching. Ensembles with
$g_{ij}^{(2)} = 1$ are referred to as fully second order coherent whereas for
bunched particles one can infer that they have the tendency to
reside together and vice versa for the antibunched
case. Figures 3(c), (d) illustrate various components of the second
order correlation function for different interaction quenches. For
the diagonal terms that refer to the middle well we observe that
$g_{MM}^{(2)} > 1$ for the whole propagation time while for the left (or right)
well we find that $g_{LL}^{(2)}$ oscillates around unity. The
latter indicates a dynamical transition from bunching to
antibunching and vice versa which has an impact also on the
$g_{MM}^{(2)}$ component. In particular, for small quenches we can see that $g_{MM}^{(2)}$ is almost unchanged
during the dynamics while $g_{LL}^{(2)}$ oscillates around unity and spends more
time below unity. This means that for small quenches we can not affect
significantly the initial distribution and two bosons are more likely
to reside in the middle well. Increasing the quench amplitude we observe that the two components are anticorrelated 
i.e. for the time intervals where $g_{LL}^{(2)}$ is smaller than unity the
corresponding component $g_{MM}^{(2)}$ for the middle well is enhanced.
Here the reduction of the $g_{LL}^{(2)}$ component is more pronounced than the
enhancement of the $g_{MM}^{(2)}$ which might indicate an impact of the initial
distribution. The off-diagonal terms $g_{LR}^{(2)}$, $g_{LM}^{(2)}$ with respect to the
left well are always lower than 1 and anticorrelated. Also, for every time during the dynamics 
$g_{LM}^{(2)}>g_{LR}^{(2)}$ holds, indicating that it is more likely for two bosons to be
one in the left and one in the middle site than one in the left and one in the right.
On the other hand, the oscillatory behaviour of $g_{ij}^{(2)}$ can again be attributed to the finite
size of our system. Concluding this part we can infer that the one-body and two-body 
correlations as shown in Figure 3 demonstrate a rich phenomenology in terms of correlation dynamics. This might pave the way for 
further investigations on how a weak density-wave 
tunneling can transport significant correlation oscillations. 

As the density-wave tunneling has been understood to lead to the ''envelope breathing'' with the character of a
breathing mode, let us now turn our attention to the study of the on-site or local breathing mode.
In general, the breathing mode then refers to a uniform expansion and
contraction of the local wavepacket. For a recent study concerning the dependence of the breathing mode frequency on the particle number as well as on the interaction strength see \cite{Schmitz}, while for further 
related and recent investigations we refer the reader to Refs \cite{Abraham3}-\cite{Abraham1}. As we shall discuss briefly here, this local breathing mode can also be triggered by 
a quench of the interaction strength in a harmonic trap. To this end, Figure 4(a) shows snapshots of the one-body density of a system consisting of two bosons in a
single harmonic trap (with ${g_{in}} = 0.05$) after an interaction
quench $\delta g = 2.8$ which mimics the dynamics within the middle well of the triple-well system. 
Here, we observe the reshaping of the density profile for different time instants as well as the formation
of oscillatory structures which indicate the existence of higher-band
effects.

Coming back to the triple well, the local breathing mode refers to
a contraction and expansion dynamics of the wavepacket in a single well,
i.e. intrawell breathing induced by an interaction quench.
In order to quantify the local breathing frequency in the
triple-well setup we define the coordinate of the center of mass of
the respective well
\begin{equation}
\label{eq:9}X_{cm}^{(i)} = \frac{{\int\limits_{{d_i}}^{{d_{i + 1}}}
{dx\left( {x - x_0^{(i)}} \right){\rho _{i}}(x)}
}}{{\int\limits_{{d_i}}^{{d_{i + 1}}} {dx{\rho _{i}}(x)} }}.
\end{equation}
Here $i=R,M,L$ stands for the right, middle and left well
respectively whereas ${x_0^{(i)}}$ refers to the middle point of the
corresponding well. On the other hand, ${d_i}$ are the coordinates of the edge points 
of an individual well and ${\rho _{i}}(x)$ the corresponding
single-particle densities. From this point of view the
preferable quantity to identify the breathing process is the
variance of the coordinate of the center of mass
\begin{equation}
\label{eq:10}var\left[ {{x^{(i)}}(t)} \right] = \sigma _{(i)}^2 =
\int\limits_{{d_i}}^{{d_{i + 1}}} {dx} {\rho _{i}}(x){\left(
{{x} - X_{cm}^{(i)}} \right)^2}.
\end{equation}
Therefore, the breathing frequency of
the middle well can be obtained from the spectrum of the second
moment $\sigma _M^2(\omega ) =
\frac{1}{\pi}\int{dt\sigma_M^2(t){e^{i\omega t}}}$. In Figure 4(b) we observe a dominant frequency, located at $\omega  \approx 60\Delta\omega$ (with $\Delta\omega=2\pi/T$ and T being the total propagation period)
which is approximately two times the trapping frequency of a harmonic approximation to a single well. This frequency depends only weakly on the 
interaction quench and it is related to the breathing frequency. There occur additional low 
frequency branches in Figure 4(b) which are related e.g. to the tunneling dynamics.

\subsubsection{The cradle mode induced by the over-barrier transport}

For a qualitative description of the cradle mode one has to rely on the
intrawell dynamics of $\delta\rho(x,t)$ for the left or right well as shown in Figure 1.
In particular, the generation of this mode is accompanied by a
direct over-barrier transport as a consequence of the interaction
quench. This results in a cradle mode which represents a dipole-like
oscillation. In the following, let us first illustrate the main
mechanism and then analyze in some detail the cradle mode.

Initially, in terms of its dominating spatial configuration our system consists
of two bosons in the middle well and two others each of them localized in one of the outer wells. Then we perform a sudden change in the
interaction strength which raises the energy as mentioned previously. As a
consequence with high probability at least one particle from the central well gains enough
energy to overcome the barrier (over-barrier transport), and directly moves to the outer wells where it performs
an inelastic collision with the preexisting particle initially
localized in the neighboring site. The two-particle collision leads to  
a cradle dynamics and to the dipole-like density oscillation as visualized in Figure
5. According to our simulations we observe significant over-barrier transport for $\delta g> 0.24$. This
process is most significant for the first few periods of the cradle motion
as for later times due to inelastic collisions in the left well the atom
looses part of its initial energy and can predominantly tunnel through the
barrier.

Therefore, the cradle mode as a localized wave-packet oscillation
can be produced via a variation in the respective interaction
strength. Moreover as already mentioned, is reminiscent of the dipole oscillation
in the one-body density evolution while a detailed analysis
demonstrates a major difference between the two. Indeed, the cradle mode which
is of two-body nature possesses two intrinsic frequencies that refer
to the center of mass and the relative frame of the harmonic
oscillator. As we prove in the appendix up to a good approximation
this can be modeled by a coherent state of the center of mass and relative coordinates. Finally, note that during
the evolution we can identify regions of bright and dark cradles
which are associated with an enhanced or reduced tunneling of the density from the
respective well (see also Figure 1).

Especially, as the cradle mode breaks the local reflection symmetry of the
one-body density in each well, we divide for a further investigation of this
mode (neglecting the breathing mode) each well into two equal parts left and right of the center with corresponding integrated densities ${\rho
_{a,1}}(t)$ and ${\rho _{a,2}}(t)$. Here the index $a$ refers to the corresponding well,
i.e. $a = L,M,R$ for the left, middle and right well respectively. In the following, we use as a measure of
the intrawell wavepacket asymmetry (referring to the cradle motion)
the quantity $\Delta {\rho _a}(t) = {\rho _{a,1}}(t) - {\rho
_{a,2}}(t)$. Furthermore, in order to investigate the impact of different
quenches on the system we compute the Fourier transformation of the
quantity $\Delta {\rho _a}(\omega ) = \frac{1}{\pi }\int {dt} \Delta
{\rho _a}(t){e^{i\omega t}}$ which will provide us with the
evolution of the frequencies of the respective modes for different
quench amplitudes. Figure 6(a) presents the resulting
frequencies from the $\Delta {\rho _L}(\omega )$ versus the
respective interaction quench $\delta g$ for 110 different quenches from
weak-to-strong interactions, where the amplitude $\delta g$ varies
from 0.04 to 4.5.

Firstly, from Figure 6(a) we can identify one dominant branch which is
insensitive to the quench amplitude and its frequency is that of the 
cradle mode. This branch corresponds to the cradle intrawell oscillation 
and will be referred to in the following as the cradle branch. A modulation of the frequency of the cradle motion 
can be achieved by tuning the barrier height, i.e. we can reduce its
frequency using lower barriers and vice versa. 

Besides the cradle branch, three interwell tunneling branches show up in the 
spectrum of $\Delta {\rho _L}(\omega )$ with a relatively weak amplitude. Among them 
we can distinguish the contribution of the highest frequency tunneling branch. The latter together with the 
branch of the cradle experience an avoided-crossing at $\delta g \simeq 2.8$ in the course 
of which both amplitudes are enhanced.

For a more detailed analysis of the above observations, let us assume that
initially the state of the system in terms of the eigenstates of the
final Hamiltonian is given by a linear superposition of the form
$\left|{\psi (0)} \right\rangle  = \sum\limits_{i;\alpha;I}
{{C_{i}^{\alpha;I}}\left| i \right\rangle _{\alpha;I}}$. Then at an arbitrary
time instant $t$ the expectation value of the intrawell asymmetry operator can
be expressed as
\begin{equation}
\label{eq:11}\left\langle \psi  \right|\Delta \widehat \rho \left|
\psi  \right\rangle  = \sum\limits_{i;\alpha;I}  {{{\left| {{C_{i}^{\alpha;I} }}
\right|}^2}{}_{I;\alpha}\left\langle i  \right|} \Delta \widehat \rho \left|
i  \right\rangle_{\alpha;I}  + 2\sum\limits_{i  \ne j }
{{\mathop{\rm Re}\nolimits} \left( {C_{i}^{\alpha;I *}{C_{j}^{\beta;I} }}
\right){}_{I;\alpha}\left\langle i  \right|} \Delta \widehat \rho \left|
j  \right\rangle_{\beta;I} \cos \left[ {\left( {{\omega _{i}^{\alpha;I} } -
{\omega _{j}^{\beta;I} }} \right)t} \right].
\end{equation}

Here, the first term refers to the average part whereas the second
term demonstrates an oscillatory behaviour. In the following, we
will concentrate on the oscillatory term of this expectation value
which essentially describes the cradle motion. As also illustrated in the analytical 
expression for the cradle mode (see Appendix) the dominant oscillation terms  ${}_{\alpha;I}\left\langle i \right|\Delta
\widehat \rho \left| j  \right\rangle_{\beta;I}  \ne 0$ are given by the eigenstates  ${\left| {i} \right\rangle _{1;0}}$ and 
${\left| {i} \right\rangle _{1;1}}$ within which ${\left| {2,1,1} \right\rangle _0}$ and  ${\left| {2,1,1} \right\rangle _1}$ significantly 
contribute respectively. Consequently, the corresponding oscillation frequency matches the energy diference between these eigenstates which is to a good approximation 
given by the energy difference ($\Delta\epsilon$) between ${\left| {2,1,1} \right\rangle _0}$ and ${\left| {2,1,1} \right\rangle _1}$. Meanwhile, ${\left| {2,1,1} \right\rangle _1}$ also contributes to 
the eigenstates ${\left| {i} \right\rangle _{2;0}}$ and ${\left| {i} \right\rangle _{3;0}}$ 
of the double pair and triple modes respectively thus leading to a non-zero ${}_{1;0}\left\langle i \right|\Delta
\widehat \rho \left| j  \right\rangle_{2(3);0}  \ne 0$, and therefore to the observed tunneling branches. The above mechanism is resonant between   
${\left| {2,1,1} \right\rangle _0}$ and ${\left| {2,1,1} \right\rangle _1}$ for a particular quench amplitude $\delta{g}$.   

To verify our statements, let us calculate the number state energy differences between the aforementioned states and compare them with the eigenenergy difference 
in the full spectrum (Figure 6(a)). In this manner we indeed find good agreement. We illustrate the $\delta{g}$-dependence of these
frequencies in Figure 6(a) with the white full dots and open circles on top of the exact
avoided-crossing implying the reliability of our above statements. Indeed, we observe only very minor deviations of numerical ML-MCTDHB results and the description via eq.(14). However, 
the intensities do differ significantly, see Figure 6(a). 
\begin{figure}[h]
        \centering
                \includegraphics[width=0.90\textwidth]{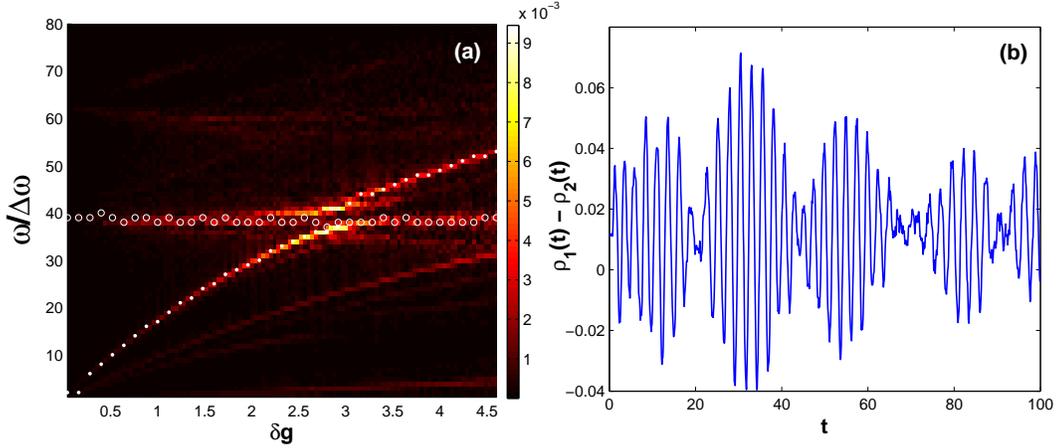}
                \caption{(a) Fourier spectrum as a function of the interaction quench of the intrawell asymmetry
                $\Delta {\rho _L}(\omega )$ for the left well (see text). The same spectrum can be obtained for the right well.
                 The frequency units $\omega/\Delta\omega$ are normalized with $\Delta\omega=2\pi/T$ and T being the respective propagation period.
                 An avoided crossing takes place between the tunneling and the cradle modes where we observe an enhancement
                 of the mode amplitudes at least for finite time propagation periods. The full white dots in the tunneling branch correspond 
                 to the intraband frequency $\Delta {\omega _1}$ between the states ${\left| {2,1,1} \right\rangle _0}$ and ${\left| {3,0,1} \right\rangle _0}$, whereas the 
                 empty circles in the branch of the cradle mode refer to the frequency
                 $\Delta {\omega _2}$ for the states ${\left| {2,1,1} \right\rangle _0}$ and ${\left| {2,1,1} \right\rangle _1}$ describing the cradle-like process (see text). As a consequence 
                 we notice the occurence of a beating (b) for the cradle in the region of the
                 avoided crossing.}
\end{figure}

In conclusion, the avoided crossing and the accompanying enhancement
indicate that the local intrawell dynamics can couple to the
interwell dynamics. In turn, this induces a resonance between the
two dynamical processes which can enhance the local and
long-range dynamics. The fact that the cradle mode can be coupled
with a mode of the interwell tunneling is remarkable. This gives rise to the possibility of controlling the 
''global dynamics'' by triggering the ''local dynamics''. Especially, we can tune the
characteristic frequency of the tunneling mode to become resonant with the
corresponding frequency of the cradle mode by means of tuning the
quench amplitude. Increasing further the
quench amplitude we can drive the system again out of resonance.

As a consequence of this avoided crossing the intrawell asymmetry $\Delta\rho$ dynamics features a
beating, as shown in Figure 6(b), which corresponds to two
dominant frequencies in the ${\rho _1}(\omega ) - {\rho _2}(\omega
)$ spectrum. Indeed,
let $\Delta {\omega _1}(\delta g) = {E_\alpha } - {E_\beta }$ be the
frequency of the respective intraband tunneling (${\left| {2,1,1}
\right\rangle _0} \to {\left| {3,0,1} \right\rangle _0}$). Assume further that 
$\Delta {\omega _2}(\delta g) = {E_\gamma } - {E_\sigma }$ refers to
a frequency of a process that includes a ground and the first excited state of
the single pair mode taking into consideration that we refer to one of the outer wells
(left or right) so we need two particles there. In this manner, there exists a
region of critical quench amplitudes $\delta {g_{cr}}$ which
corresponds to the avoided crossing where ${\Delta\omega _1}(\delta
{g_{cr}}) \approx {\Delta\omega _2}(\delta {g_{cr}})$ and the system
features a degeneracy between the states ${\left| {3,0,1}
\right\rangle _0}$ and ${\left| {2,1,1} \right\rangle _1}$. 

From the above discussion, one can infer that
a representative wavefunction describing the cradle process in terms
of Fock states for the left well (and similarly for the right) can
in principle be written as
\begin{equation}
\label{eq:12}\left| \varphi  \right\rangle _L^{cradle} =
{{C_0}(\delta g,t)}{\left| {2,1,1} \right\rangle _0} +
{{C_1}(t)}{\left| {2,1,1} \right\rangle _1} \equiv \left| {2,1,1}
\right\rangle _L^D,
\end{equation}
where the coefficients $C_{0}$ and $C_{1}$ denote the probability
amplitudes for the corresponding state. Note also that the amplitude
of the zeroth state ${\left| {2,1,1} \right\rangle _0}$ depends on
the quench while the one for the first-excited state ${\left| {2,1,1}
\right\rangle _1}$ is essentially constant, i.e. independent of $\delta{g}$.

Taking advantage of the previous description we can construct an
effective Hamiltonian for this process. Thus, if we denote by
$\left\{ {\left| {{{\vec n }_0}} \right\rangle } \right\}$ the
corresponding truncated basis vectors, the effective Hamiltonian
obtained from (1) in this subspace will be of the form
\begin{equation}
\label{eq:13}{H_{\text{eff}}} = \sum\limits_{{{\vec n }_0}} {{\epsilon
_{{{\vec n }_0}}}} \left| {{{\vec n }_0}} \right\rangle \left\langle
{{{\vec n }_0}} \right| + \sum\limits_{{{\vec n }_0},{{\vec m }_0}}
{{J_{{{\vec n }_0},{{\vec m }_0}}}} \left| {{{\vec n }_0}}
\right\rangle \left\langle {{{\vec m }_0}} \right|,
\end{equation}
where ${J_{{{\vec n }_0},{{\vec m }_0}}} = \left\langle {{{\vec n
}_0}} \right|{H_{\text{eff}}}\left| {{{\vec m }_0}} \right\rangle$ is the
effective tunneling amplitude and ${\epsilon _{{{\vec n }_0}}} = \left\langle {{{\vec n }_0}} \right|{H_{\text{eff}}}\left|
{{{\vec n }_0}}\right\rangle$.

Therefore the representative subspace providing the mode-coupling within a minimal model 
consists of the states $ {\left| {2,1,1} \right\rangle }_0, {\left| {3,0,1} \right\rangle}_0, {\left| {2,1,1} \right\rangle }_1$. In terms of the
corresponding effective Hamiltonian the respective term for the cradle
mode is $|2,1,1\rangle_{1~0}\langle 2,1,1|$ whereas the
term $|2,1,1\rangle_{0~0}\langle{3,0,1}|$ reflects the
tunneling process. Thus from the Hamiltonian (16) one can realize a
three-level system consisting of the states according to their
energetical order: ${\left| {2,1,1} \right\rangle _0} \equiv \left|
1 \right\rangle $, ${\left| {3,0,1} \right\rangle _0} \equiv \left|
2 \right\rangle $, ${\left| {2,1,1} \right\rangle _1} \equiv \left|
3\right\rangle$. In this manner, we take into account an energy detuning $\Delta
$ between the states $\left| 2 \right\rangle $ and $\left| 3
\right\rangle $ whereas due to the fact that the level $\left| 1
\right\rangle $ is weakly coupled with the other levels we 
neglect the respective tunneling amplitudes, i.e. ${J_{12}} = {J_{13}} = 0$.
Therefore, we can reduce our problem to a two level system
realizing the Hamiltonian ${H_{\text{eff}}} = \sum\limits_{i = 1}^3
{{\varepsilon _i}\left| i \right\rangle \left\langle i \right|} +
{J_{23}}\left| 2 \right\rangle \left\langle 3 \right| + h.c$ which
is known to exhibit an avoided level crossing and can be solved
analytically.

\begin{figure}[h]
        \centering
                \includegraphics[width=0.90\textwidth]{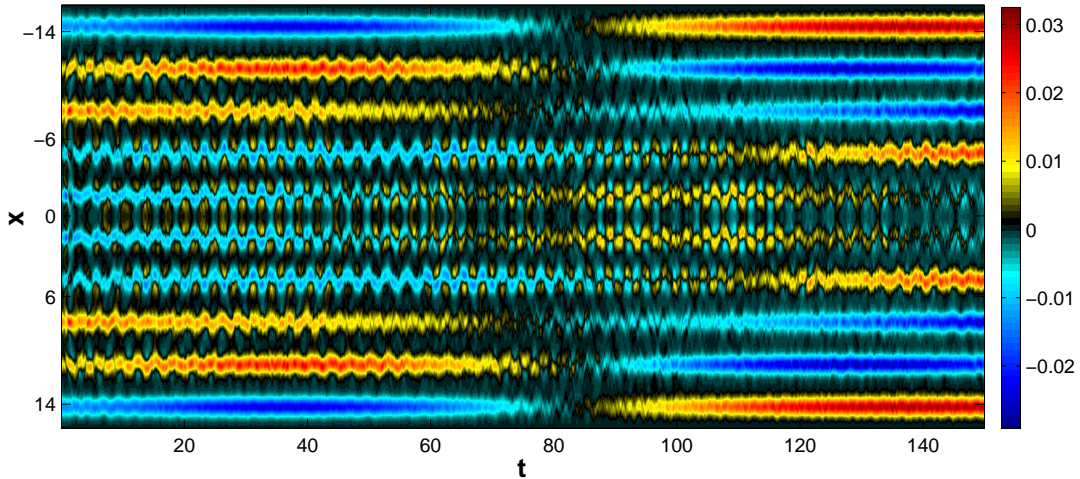}
                \caption{The fluctuations $\delta\rho(x,t)$ of the one-body density for five atoms in ten wells.
                Initially we observe the emergence of the over-barrier transport and subsequently 
                the cradle and the tunneling modes. The setup is initially prepared in the superfluid ground state
                with $g_{in}=0.05$ and is suddenly quenched with 
                $\delta g = 4.0$.}
\end{figure}
In the remaining part of this section, we proceed to the investigation of a system with
filling factor $\nu  < 1$ in order to generalize our findings. More
precisely, among others we demonstrate that the cradle mode can
also be found in the inner-well dynamics for a setup with ten
wells, which reveals in particular that it is independent of the employed hard-wall
boundaries.

\subsection{Filling factor $\nu  < 1$}

Let us consider here the case of five bosons in a ten-well finite lattice.
Concerning the ground state analysis with filling factor $\nu<1$,
the most important aspect is the spatial redistribution of the
atoms as the interaction strength increases. The non-interacting ground state (g=0) is the product of the single-particle eigenstates spreading across 
the entire lattice, while due to the hard-wall boundary conditions the two central wells of the potential are slightly more populated. As the repulsion increases within the weak interaction regime the atoms are pushed to the outer sites which gain and lose population 
in the course of increasing $g$, while the particle number fluctuations are more pronounced for
the wells with a lower population \cite{Brouzos}.
\begin{figure}[h]
        \centering
                \includegraphics[width=0.90\textwidth]{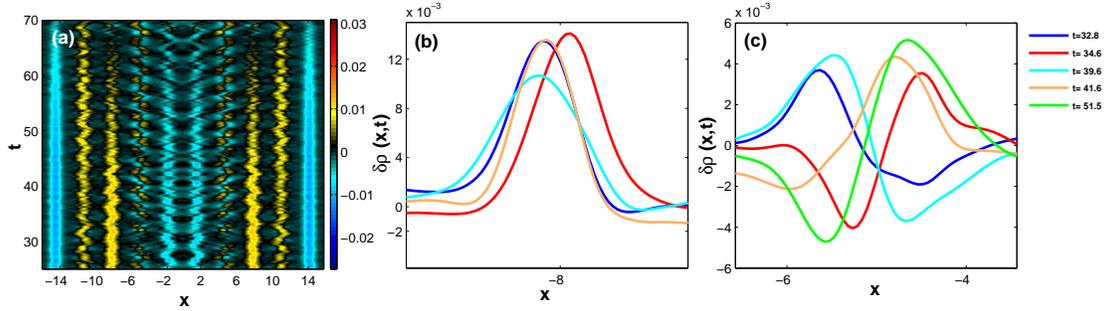}
                \caption{The response of the fluctuating part $\delta\rho(x,t)$ of the one-body density after an abrupt change in the
                inter-particle repulsion. The setup consists of five particles in ten wells and the initial state corresponds to
                interaction strength $g_{in}=0.05$. Shown are $\delta\rho(x,t)$ for (a) a given time period of the
                cradle-like process for $\delta{g}=2.4$ and the respective fluctuating one-body density profiles at different time instants for (b)
                the second well indicating the ground state of the cradle state and (c) the fourth well demonstrating the first-excited state of the
                cradle process. The nodal structure indicates the occupation of excited Wannier states in the respective
                well whereas the oscillatory behaviour visualizes the cradle process.}
\end{figure}
It is also important to notice that in such a setup the one-body density will not become uniform
even for strong interactions. In addition, the particle number fluctuations
saturate to a relatively high value (for $g \approx 3.5$) in
accordance with the existence of the delocalized phase. Also, in such a case of incommensurability
due to the delocalized fraction of particles the long-range one-particle correlations do not vanish even for strong interactions.

\begin{figure}[h]
        \centering
                \includegraphics[width=0.90\textwidth]{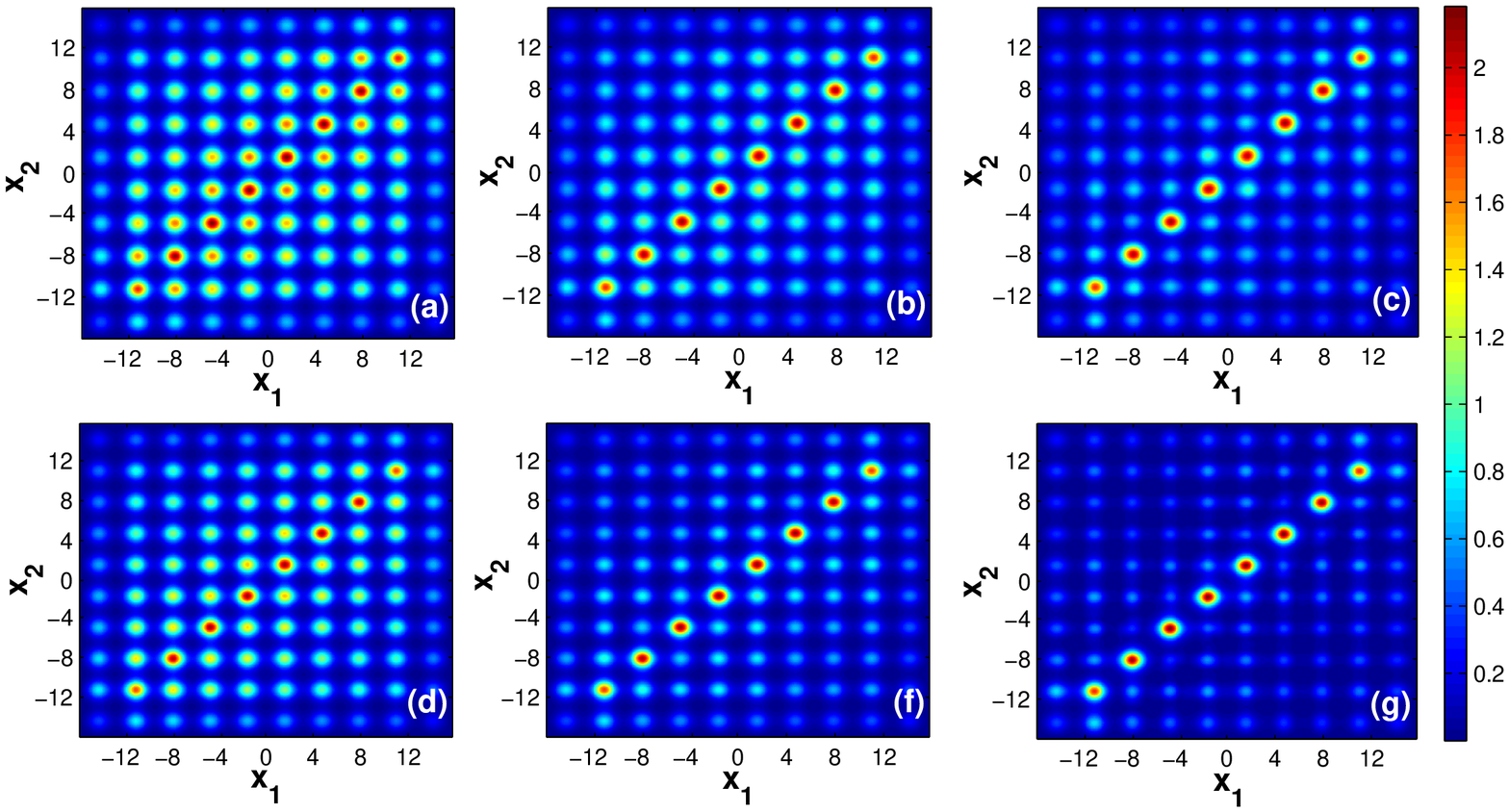}
                \caption{Off-diagonal one-body reduced density matrix ${\rho ^{(1)}}(x,x';t_{1})$ for five
                particles in 10 wells for two different time instants $t_{1}=7.0$ (a,b,c) and $t_{2}=42.8$ (e,f,g) during the evolution. Shown are
                three values of the interaction quench (a,d) $\delta g = 1.4$, (b,f) $\delta g = 2.6$ and (c,g) $\delta g =3.8$.
                The elimination of the off-diagonal spots for strong quench amplitudes indicates the difference in the tunneling
                process. }
\end{figure}
In the following, we explore the dynamics following a sudden interaction quench at time
t=0 which is applied to the ground state in the weak interaction
regime, $g_{in}=0.05$. Figure 7 demonstrates the response of the system on the one-body level namely $\delta\rho(x,t)$ after
a strong interaction quench $\delta{g}=4.0$, from
which we can easily identify the emergence of three modes. 
Initially we observe an over-barrier transport and then the cradle
and the tunneling modes. The lattice symmetry (even number of wells)
leads to the lack of the local breathing mode. Concerning the cradle mode,
this would be a superposition of the states
$\ket{1,1,2,0,0,0,0,...}_{0}+permut$ and
$\ket{1,1,2,0,0,0,0,...}_{i}+permut$ with $i \ne 0$, where $permut$
stands for the spatial permutation of the occupations inside the
ket vector. Here, one can consider the minimal subspace consisting of the
above states with $i=1$ in order to proceed in an effective approach
as we did for the case $\nu>1$.

Additionally, in order to visualize the cradle process from a
one-body perspective we demonstrate in Figure 8 for a specific
quench $\delta{g}= 3.4$ the density 
fluctuations $\delta\rho(x,t)$ for different time instants. Figure
8(a) illustrates the evolution of the fluctuations for a specific
time interval following up on the over-barrier transport, where we can observe
the cradle process in each well. Subsequently, in Figures 8(b) and
8(c) we show intersections of $\delta\rho(x,t)$ focussing on
the second and the fourth wells in order to visualize the
higher-band contributions to the mode. Indeed, in Figure 8(b) we
observe that the profile of the fluctuations corresponds to almost
localized wave-packets inside the well which are spatially shifted with 
time. This process demonstrates the motion of the cradle and
corresponds to the ground state of this mode. However, Figure 8(c)
illustrates the same profile $\delta\rho(x,t)$ but for a different
well, where the appearance of at least one node indicates the
occupation of the first excited Wannier state in the well.
This behaviour together with the shift of the wave-packet indicates
the first-excited state of the cradle mode. 

Also, in the Fourier spectrum we can find the avoided-crossing between the cradle
and the tunneling frequency where the critical region of quench
amplitudes is $\delta {g_{cr}} =4.2-4.3 $ with equal cradle
frequency as for the triple well case due to the same barrier height
${V_0} = 4.5$. In turn, the avoided crossing here, if we refer to the
third well, can be explained from the dominant number states ${\left|
{1,1,2,0,...} \right\rangle _0}$, ${\left| {1,1,2,0,...}
\right\rangle _1}$ for the cradle and ${\left| {1,1,2,0,...}
\right\rangle _0}$, ${\left| {1,0,3,0,...} \right\rangle _0}$ for
the tunneling process. Therefore, we can conclude that by tuning the
interaction quench we can again realize a resonance between the
aforementioned modes.

An important observation is that as we increase
the interaction quench the tunneling process can be altered. Indeed,
this behaviour can be attributed to the fact that the higher the
quench amplitude, the larger the energy of the system becomes.
From this point of view we expect a strong link of 
the change of the spatial distribution of the atoms in the
lattice and the quench amplitude. The above behaviour
is a main characteristic of setups with filling factor $\nu  < 1$
where on-site effects are not manifested. In order to quantify 
our arguments concerning the spatial redistribution process we will rely on an analysis of the
one-body reduced density matrix ${\rho
^{(1)}}(x,x';t)$ of the dynamics provided 
by ML-MCTDHB. Its off-diagonal parts can be used as a measure of the coherence as
they indicate off-diagonal long-range order in an infinite
lattice. Although, for our finite setups we cannot conclude upon true
off-diagonal long-range order this term refers to the appearance of
short and long range one-body correlations. Thus, this quantity can
be directly linked to the tunneling process.
 In Figure 9, we depict the one-body density matrix for three different
quenches namely $\delta g = 1.4,2.6,3.8$ at two different time
instants $t_{1} = 7.0$ (a,b,c) and $t_{2}= 42.8$ (e,f,g) of the
propagation in order to indicate the change in the tunneling
process. The off-diagonal contributions fade out with
increasing quench amplitude and a tendency for concentration close to
the diagonal is observed at equal times which implies an alteration of the character of the tunneling process. In
addition, the off-diagonal part cannot vanish completely even for
strong quenches since the particles always remain delocalized. This
is a main characteristic of incommensurate setups.

 Going beyond the above examined finite setups, one can suggest a generalization for the
 wavefunction of the cradle state for a many-body system. Let $N$ be the number of sites
and $n$ the total number of bosons. Then the corresponding
generalized number state would be of the form ${\left|
{{n_1},{n_2},...,{n_N}} \right\rangle _i}$ whereas the minimum
representative wavefunction for the cradle that refers to the first well can be written as
\begin{equation}
\label{eq:15}{\left| \psi  \right\rangle ^{cradle}} =
{d_1}(t){\left| {2,{n_2},...,{n_N}} \right\rangle _0} +
{d_2}(t){\left| {2,{n_2},...,{n_{N}}} \right\rangle _1},
\end{equation}
where ${n_2} + {n_3} + ... + {n_N} = n - 2$ with $n > 2$ and $d_{1}$, $d_{2}$ denotes the amplitudes for each contribution in the above expansion. An
additional constraint is that for an even number of sites $N$ this
relation holds for all permutations, while for odd $N$ the
permutation that corresponds to a state with two particles in the
middle well indicates the presence of the local breathing motion. The extension to cradles 
in the remaining wells is straightforward.

 \section{Summary and Conclusions }

We have explored the influence of sudden interaction quenches on small bosonic ensembles in finite one-dimensional (1D) 
multi-well traps. In particular, we have mainly focussed on setups with incommensurate filling factors in order to avoid the suppression of tunneling due 
to MI phases for strong interactions. Starting from the superfluid regime, we change abruptly the effective coupling strength from weak to strong interactions. 
In this manner, we observe the emergence of tunneling, breathing and cradle processes.
Furthermore for the explanation of the dynamical behaviour of our system in terms of a band structure we employ the concept of generalized multi-band Wannier number states which are meaningful for sufficiently large barrier heights ${V_0}$. 
Although these have been constructed numerically, such a treatment is valid even in the strong interaction regime where perturbative methods fail. 

The density-wave tunneling has been linked to an effective breathing of the ''global wavepacket'' that refers to the instantaneous density distribution of the trap.
The local breathing mode has been identified as an expansion and contraction dynamics of bosons in the individual wells. Moreover, in terms of a number state analysis of the observed dynamics it is necessary to include 
higher-band contributions to describe it.
On the other hand, the cradle process, as we have pointed out, exists in almost every site of the lattice and refers to a localized wave-packet oscillation.
This mode is a consequence of the initial over-barrier transport of the particles from the central well to the outer ones due to the sudden import of energy into the system and the consequent inelastic collisions with the respective atoms 
in the outer sites. 
Therefore, we can describe this process via the coherent states of the harmonic oscillator refering here to the center of mass and the relative coordinate (see Appendix). 
The aforementioned modes are always accompanied by a tunneling process which is mainly a lowest-band phenomenon. 
During the dynamical process, regions of density dips or dark cradles in the outer sites are accompanied by enhanced breathing dynamics on the 
middle site. Each of the above modes possesses different characteristic frequencies. 
In particular, we show that one can tune the frequency of the highest branch of the tunneling mode in resonance with the frequency of the cradle mode by varying the quench amplitude. 
In turn, this resonance is associated with an avoided crossing in the frequency spectrum of these modes resulting in an enhancement of both of them. 
In this case, the system features a dominant beating.

We have computed, the dominant Fock states in the expectation value of the asymmetry operator in order to describe the dynamics associated with the avoided-crossing. 
In this manner, we have found a representative cradle state which is a superposition of the first two bands, as well as the number state most responsible for the tunneling mode that couples with the cradle in the avoided crossing.

There are at least two ways that one might pursue as a follow-up direction. 
A first possible extension is to consider smooth time-dependent interaction quenches in order to unravel the behaviour of the system in such a non-equilibrium continuously driven case or to find similarities with the so-called Kibble-Zurek mechanism \cite{Kibble}-\cite{Damski1}. 
A second path in this direction would be the study of mixtures consisting of different bosonic species.

\section*{appendix: Remarks on the Cradle state.}

In this Appendix we will briefly discuss the derivation of the cradle state. 
This state, as we have already mentioned in the main text, refers to an oscillation of two wavepackets of minimal uncertainty in a single well which we model as a harmonic trap. 
The creation of the two-particle cradle state in a single well corresponds to the collision between a particle injected to the
well with another particle which is initially localized in the
minimum of the well. We further model each particle as a localized
one, where the first one is centered in the trap and the other one is
displaced from the minimum by an amount ${x_0}$. In this manner,
taking advantage of the exactly solvable model of the harmonic
oscillator, we can derive the initial wavefunction of the cradle
state. In the following, we will use the natural units $m = \hbar  =
1$. Due to the harmonic potential approximation
we can separate the motion into the relative ${X_r} = \left(
{{x_1} - {x_2}} \right)/\sqrt 2 $ and center-of-mass ${X_c} =
\left( {{x_1} + {x_2}} \right)/\sqrt 2 $ coordinates. Adopting these coordinates
the initial wavefunction reads
\begin{equation}
\label{eq:16}{\psi _0}({X_c},{X_r};0) = {\left( {\frac{\omega }{\pi
}} \right)^{1/2}}{e^{ - \frac{\omega }{2}{{\left( {{X_c} -
\frac{x_{0}}{{\sqrt 2 }}} \right)}^2}}}({e^{ - \frac{\omega
}{2}{{\left( {{X_r} + \frac{x_{0}}{{\sqrt 2 }}} \right)}^2}}} + {e^{
- \frac{\omega }{2}{{\left( {{X_r} - \frac{x_{0}}{{\sqrt 2 }}}
\right)}^2}}}).
\end{equation}
 Thus, the intrawell oscillation can be separated into two parts: a) the
center-of-mass motion which is an effective one-body problem and b) the
relative motion that refers to a reduced two-body problem. Therefore, we can
easily show that the wavefunction of the center-of-mass at any time
$t > 0$ is described by
\begin{equation}
\label{eq:17}\psi ({{\rm X}_c};t) = \frac{1}{{\sqrt {\sqrt \pi}
}}{e^{ - \frac{1}{2}({X_c} - {x_0}\cos \omega t) - i(\frac{{\omega
t}}{2} + {X_c}{x_0}\sin \omega t)}},
\end{equation}
which is the well-known coherent state solution of the harmonic
oscillator. This wave-packet oscillates around the minimum of the
potential in a simple harmonic trap without changing its shape while
it satisfies the minimum-uncertainty, i.e. $\Delta p\Delta x = \hbar /2$.
On the other hand, the corresponding wavefunction of the relative
frame reads
\begin{equation}
\label{eq:18}\psi ({{\rm X}_r};t) = \sum\limits_n {{C_{2n}}{e^{i{\omega _{2n}}t}}{\varphi _{2n}}},
\end{equation}
where ${{\varphi _{2n}}}$ are the even eigenstates of the trapping
potential $V(x) = \frac{1}{2}m\omega^{2} {x^2} + g\delta (x)$. These
eigenfunctions are known as the Weber functions. Thus, we can conclude that
the cradle state contains two characteristic frequencies: a) the
frequency ${\omega _C} = \omega $ that refers to the motion of the
center-of-mass and b) the frequency ${\omega _r} = {\omega _{2n +
2}} - {\omega _{2n}} \approx 2\omega $ of the relative frame.
The above comments lead us to the conclusion that the major
difference between the cradle state that we have found here and the
dipole state of a many-body system is that our state contains the
two frequencies ${\omega _C}$ and
${\omega_r}$ while the many-body collective state has just one.

Comparing the analytical results with the exact numerical ones obtained from 
the ML-MCTDHB method we conclude that in our case we observe only the
center-of-mass oscillation in the frequency spectrum. This is a
consequence of the fact that the quantity $\Delta \rho  = {\rho _1}
- {\rho _2}$ that we have used to measure the intrawell asymmetry
can describe only the motion of the center-of-mass.

\section*{Acknowledgments}

L.C. and P.S. gratefully acknowledge
funding by the Deutsche Forschungsgemeinschaft (DFG) in
the framework of the SFB 925 “Light induced dynamics and
control of correlated quantum systems. The authors would like to thank P. 
Giannakeas, C. Morfonios and S. Kr$\ddot{o}$nke for fruitful discussions. We also thank J.
Stockhofe for a careful reading of the manuscript.

{}

\end{document}